# Detecting and Characterizing Propagation of Security Weaknesses in Puppet-based Infrastructure Management

Akond Rahman, *Member, IEEE* and Chris Parnin *Member, IEEE*

**Abstract**—Despite being beneficial for managing computing infrastructure automatically, Puppet manifests are susceptible to security weaknesses, e.g., hard-coded secrets and use of weak cryptography algorithms. Adequate mitigation of security weaknesses in Puppet manifests is thus necessary to secure computing infrastructure that are managed with Puppet manifests. A characterization of how security weaknesses propagate and affect Puppet-based infrastructure management, can inform practitioners on the relevance of the detected security weaknesses, as well as help them take necessary actions for mitigation. To that end, we conduct an empirical study with 17,629 Puppet manifests mined from 336 open source repositories. We construct **Taint** Tracker for **Pup**pet Manifests (**TaintPup**), for which we observe 2.4 times more precision compared to that of a state-of-the-art security static analysis tool. TaintPup leverages Puppet-specific information flow analysis using which we characterize propagation of security weaknesses. From our empirical study, we observe security weaknesses to propagate into 4,457 resources, i.e, Puppet-specific code elements used to manage infrastructure. A single instance of a security weakness can propagate into as many as 35 distinct resources. We observe security weaknesses to propagate into 7 categories of resources, which include resources used to manage continuous integration servers and network controllers. According to our survey with 24 practitioners, propagation of security weaknesses into data storage-related resources is rated to have the most severe impact for Puppet-based infrastructure management.

## 1 INTRODUCTION

INFRASTRCTURE as code (IaC) is the practice of automatically managing computing infrastructure, such as continuous integration servers, production web servers, load balancers, or data storage, typically provisioned on public cloud services [29]. Use of IaC languages, such as Puppet has yielded benefits for information technology (IT) organizations. For example, Ambit Energy, an energy distribution company, increased their deployment frequency by a factor of 1,200 using Puppet [49]. KPN, a Dutch telecommunications company, uses Puppet manifests to manage its 10,000 servers [50]. Use of Puppet helped KPN in regulatory compliance and faster resolution of customer service requests [50].

Despite reported benefits, Puppet manifests can contain security weaknesses [53], [55], which can leave computing infrastructure susceptible to large-scale security attacks. In recent years, security weaknesses that appear in Puppet manifests, such as hard-coded passwords and use of weak cryptography algorithms, have been a contributing factor in multiple high-profile security incidents. For example, hard-coded passwords were leveraged to gain unauthorized access to Uber's servers, which resulted in data exposure for 57 million customers and 600,000 Uber drivers [41], [58]. In another incident, use of weak encryption algorithms for Amazon S3 data storage allowed malicious attacks to access over a billion health records [19], [20].

The above-mentioned examples showcase the need for proactively detecting security weaknesses in Puppet manifests. SLIC, a state-of-the-art security static analysis tool for Puppet, based on pattern matching [13], [53], [55] can be used to detect security weaknesses in Puppet manifests. Unfortunately, SLIC can be prone to reporting false positives [11], which deter practitioners from adopting or taking actions [17], [32], [56]. For example, Bhuiyan and Rahman [11] found that SLIC to generate 1,560 false positives for 2,764 Puppet manifests. Generation of such false positives would render security weakness detection impractical for practitioners, leaving security weaknesses unmitigated in Puppet manifests. Thus, enhanced static analysis tools are required for detecting security weaknesses in Puppet manifests.

Examples from the open source software (OSS) domain provide clues on how Puppet-related security static analysis can be enhanced. Let us consider two Puppet manifests from an OSS repository [37] that use SHA1, a weak encryption algorithm. According to the Common Weakness Enumeration (CWE), use of weak encryption algorithms, such as SHA1 is "*dangerous because a determined attacker may be able to break the algorithm and compromise whatever data has been protected*" [42]. In one manifest (Figure 1a), the weakly encrypted password propagates into a Puppet resource [35] used for storing user authentication, potentially allowing malicious users to access the server. In the second manifest (Figure 1b), although a SHA1 hash is created, it never propagates into a resource to manage any relevant computing infrastructure. From these examples, we can see that a security static analysis tool for Puppet must first detect potential security weaknesses, and then determine propagation into the underlying resources used to manage relevant computing infrastructure. This second step is crucial, as, not only does it help eliminate false positives (e.g., the weakness in Figure 1b), but information about propagation and associated resources can also help in determining relevance of detected weaknesses [61].

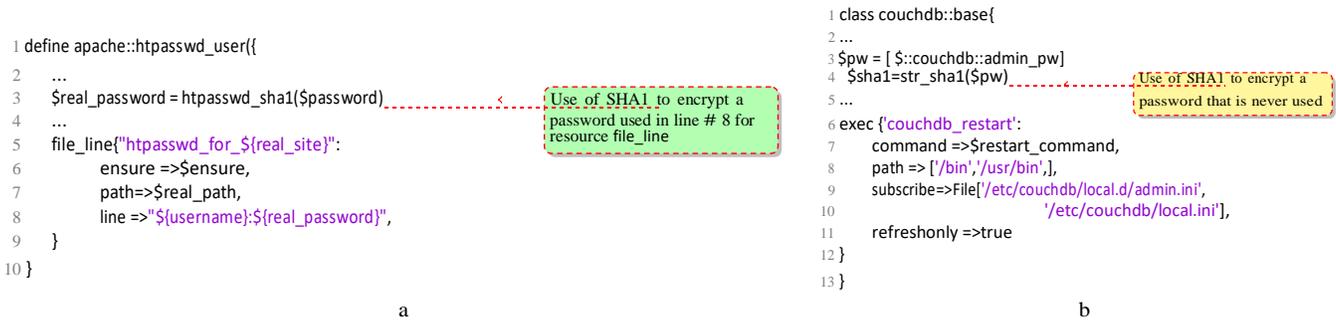

Fig. 1: OSS Puppet manifests that use SHA1, a weak encryption algorithm. In one manifest (Figure 1a), the hash propagates into a resource to setup a password file. In the second manifest (Figure 1b), although a SHA1 hash is created, it never propagates into any resources used to manage computing infrastructure.

Puppet uses a state-based approach for infrastructure management [35], which necessitates development of novel static analysis tools for detection of security weakness propagation. Puppet infers the desired infrastructure state is from the Puppet manifest with code constructs, such as resources [35]. Puppet will identify the differences between the existing and desired infrastructure states, and only apply changes if there are differences between desired and infrastructure states. Along with applying a state-based infrastructure management, Puppet allows multiple categories of information flows with code constructs, such as nodes, modules, and resources [35]. Not all of these code constructs are used to manage infrastructure. Tracking all information flows will not only be computation intensive, but also lead to generating false positives. Therefore, for detecting security weakness propagation a static analysis tool must separate and track information flows that are only used for managing infrastructure.

In this paper, we construct **Taint** Tracker for **Pup**pet Manifests (***TaintPup***), which applies Puppet-specific information flow analysis that helps us to detect and understand how security weaknesses propagate into infrastructure managed with Puppet resources. TaintPup leverages resources, i.e., code elements that are pivotal to account for state-driven infrastructure management, along with the corresponding information flows. With TaintPup, we conduct an empirical study with 17,629 Puppet manifests mined from 336 OSS repositories. We quantify how propagation detection improves identification of security weaknesses. Next, we investigate the categories of resources into which security weaknesses propagate. Finally, we survey 24 practitioners, and observe their perceptions for the identified resource categories. Dataset and source code used in our paper is available online [51].

Specifically, we answer the following research questions:

- **RQ$_1$**: *How does propagation detection improve security weakness identification in Puppet manifests?*
- **RQ$_2$**: *How frequently do security weaknesses propagate into resources?*
- **RQ$_3$**: *What are the resource categories into which security weaknesses propagate?*
- **RQ$_4$**: *What are the practitioner perceptions of the identified resources into which security weaknesses propagate?*

**Contributions**: We list our contributions as follows:

- A categorization of resources into which security weaknesses propagate;
- An analysis of how frequently security weaknesses propagate into resources;
- An evaluation of practitioner perceptions for the identified categories of resources into which security weaknesses propagate; and
- A static analysis tool called TAINTPUP that identifies the resources into which security weaknesses propagate.

## 2 TAINT TRACKER FOR PUPPET: TAINTPUP

In this section, first, we describe the construction of TaintPup. Next, we provide the methodology and the answer to $RQ_1$.

### 2.1 Construction of TaintPup

We use this section to describe TaintPup's construction. TaintPup's construction is informed by accounting the following properties unique to Puppet:

- ***State-driven Infrastructure Management***: Puppet uses a state-driven approach where manifests are developed in a manner so that it reaches a desired state [35]. During execution *first* Puppet will infer what is the desired infrastructure state from the Puppet manifest. *Second*, Puppet will identify the differences between the existing and desired infrastructure states, and only apply changes if there are differences between desired and infrastructure states. A tool that aims to detect security weakness propagation must account for Puppet's state-driven approach for infrastructure management. To address this challenge, we construct data dependence graphs, where we track if a security weakness propagates, and affects infrastructure management as described in Section 2.1.3.
- ***Infrastructure-oriented Information Flow***: Puppet allows for multiple categories of information flows with code constructs, such as nodes, classes, modules, and resources [35]. However, not all of these code constructs are used to manage infrastructure. Tracking all information flows will not only be resource intensive, but also lead to generating false positives. To address this challenge we perform three activities:

- <u>Syntax Analysis</u>: As described in Section 2.1.1, TaintPup performs syntax analysis by applying code element extraction, expression classification, and membership preservation of attributes.
- <u>Security Weakness Identification</u>: As described in Section 2.1.2, TaintPup applies rule matching to limit the scope of the information flows that need to tracked.
- <u>Taint Tracking via Data Dependence Graph</u>: As described in Section 2.1.3, TaintPup track the information flows for code elements that constitute a security weakness with data dependence graphs. In this manner, TaintPup only reports a security weakness if that security weakness is being used by a resource.

### 2.1.1 Syntax Analysis

Till date, SLIC [53] is the state-of-art security static analysis tool for Puppet [13]. While Rahman et al. [53] reported SLIC to have an average precision of 0.99, Bhuiyan and Rahman [11] found SLIC's precision to be as low as 0.72. Furthermore, they [11] reported SLIC to generate 1,560 false positives for 2,764 Puppet scripts. Reasons for false positives generated by SLIC can be attributed to two types of parsing-related limitations [53]: *(i) code element parsing:* while parsing code elements, SLIC generates false positives. For example, SLIC identifies $admin_password = pick($access_hash['password']) as a hard-coded secret, even though a function (pick()) is used; and *(ii) value parsing:* SLIC has limitations in parsing values assigned to Puppet variables. For example, SLIC fails to identify db_admin_password=undef as a false positive because it parses undef as a string. undef in Puppet is not a string, and is actually equivalent to that of NIL in Ruby [35]. TaintPup applies the following syntax analysis to account for the above-mentioned limitations of SLIC.

**Code Element Extraction**: *First*, TaintPup uses 'puppet parser dump (PPD)' [35] to identify non-comment code elements in a Puppet manifest. The benefit of using PPD is that with PPD output, TaintPup does not have to apply heuristics for code element extraction, contrary to SLIC [53]. PPD does not provide any API methods, which necessitates additional pre-processing. PPD converts a Puppet manifest to a single string of tokens where the types of each token is identified. Using stack-based parsing [7] TaintPup extracts tokens and their types from the PPD output. For example, PPD will parse $db_user='dbadmin' as (=($db_user 'dbadmin')), which in turn will be used by TaintPup to determine that $db_user is a variable, and the value 'dbadmin' is assigned to the variable. *Second*, TaintPup applies pattern matching to detect comments.

Using this step, TaintPup extracts attributes, resources, and variables. Each identified attribute is provided an unique identifier based on the resource and the manifest it appears in.

**Expression Classification**: In Puppet, an expression is an attribute or a variable to which a value is assigned directly, or indirectly, e.g., via a variable or a function [35]. TaintPup identifies three types of expressions: string, function, and parameter. An attribute or a variable that is directly assigned a string value is called a string expression. An attribute or a variable that is assigned a value using a function call is a function expression. A variable that is used as a class parameter, and is directly assigned a string value is called a parameter expression. For example, $db_user='dbadmin' is a string expression, $admin_password = pick($access_hash['password'])) is a function expression, and $workers='1' in class($workers='1') .{. }s a parameter expression. By identifying these expressions TaintPup mitigates SLIC's limitations related to code element and value parsing.

**Membership Preservation of Attributes**: A single Puppet manifest can contain resources, where each resource can have $>= 1$ attributes [35]. Furthermore, an attribute with the same name can appear for multiple resources [35]. TaintPup uses hash maps to map an attribute to its corresponding resource and manifest. These hash maps are later used to construct data dependence graphs (DDGs) that is discussed in Section 2.1.3. Figure 2 shows how the three attributes presented in Figure 1a ('ensure', 'path', 'line') are mapped to their resource file_line.

### 2.1.2 Security Weakness Identification

*First, TaintPup applies rule matching* on extracted function expressions, parameter expressions, string expressions, and resources from Section 2.1.1 to identify six security weakness categories. The rules used by TaintPup are listed in Table 1, where we provide the names, definitions, and rules for each security weakness category. Patterns used by the rules are listed in Table 2. All rules and patterns are provided by Rahman et al. [53]. *Second, TaintPup identifies variables*, which are used in any security weakness that belongs to any of the following categories: admin by default, empty password, hard-coded secret, invalid IP address binding, use of HTTP without TLS, and use of weak cryptography algorithms. TaintPup also keeps track of attributes for which a security weakness appears.

### 2.1.3 Taint Tracking via Data Dependence Graph

TaintPup applies taint tracking by constructing DDGs similar to prior research [39]. A DDG is a directed graph, which consists of a set of nodes and a set of edges. DDGs used by TaintPup applies information flow analysis by identifying def-use relationships [7].

Def-use relationships leverage the definition of reachability [7]. A weakness $x$ reaches another attribute or variable $y$, if $y$ uses $x$ and there are no other code elements between $x$ and $y$ that changes the value of $x$. We use Table 3 to demonstrate reachability. In the first row, $magnum_proto ="http:" is a string expression with the

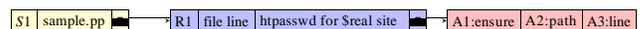

Fig. 2: An example of how TaintPup maps attributes listed in Figure 1 to their resource (file_line).

TABLE 1: Rules to Detect Security Weaknesses

| Category | Definition | Rule |
|---|---|---|
| Admin by default | Administrative privileges for users by default [53] | $(isParameter(x)) \land (isAdmin(x.name) \land isUser(x.name))$ |
| Empty password | Using a string of length zero for a password [53] | $(isAttribute(x) \lor isVariable(x)) \land ((length(x.value) == 0 \land isPassword(x.name))$ |
| Hard-coded secret | Revealing sensitive information (user names, passwords, private keys) [53] | $(isAttribute(x) \lor isVariable(x)) \land (isUser(x.name) \lor isPassword(x.name) \lor isPvtKey(x.name)) \land (length(x.value) > 0)$ |
| Invalid IP address binding | Assigning '0.0.0.0' as an IP address [53] | $((isVariable(x) \lor isAttribute(x)) \land (isInvalidBind(x.value))$ |
| Use of HTTP without TLS | Using HTTP without TLS [53] | $(isAttribute(x) \lor isVariable(x)) \land (isHTTP(x.value))$ |
| Use of weak crypto. algo. | Using MD5 and SHA1 [53] | $(isFunction(x) \land usesWeakAlgo(x.name))$ |

TABLE 2: Patterns Used by Rules in Table 1

| Function | String Pattern |
|---|---|
| *isAdmin*() [53] | 'admin' |
| *isHTTP* () [53] | 'http:' |
| *isInvalidBind*() [53] | '0.0.0.0' |
| *isPassword*() [53] | 'pwd', 'pass', 'password' |
| *isPvtKey*() [53] | '[pvt/priv]+*[cert/key/rsa/secret/ssl]+' |
| *isUser*() [53] | 'user' |
| *usesWeakAlgo*() [53] | 'md5', 'sha1' |

TABLE 3: An Example to Demonstrate Reachability

| Coding Pattern | Reachability |
|---|---|
| 1. $magnum_proto = ''http:'' <br> 2. package{ 'sample' <br>   ensure => '4.2.1-5.fc25', <br>   url =>  $magnum_proto  ''localhost:8888'' <br> } | $magnum_proto = ''http:'' reaches url |
| 1. $magnum_proto = ''http:'' <br> 2. $magnum_proto = ''ftp:'' <br> 3. package{ 'sample' <br>   ensure => '4.2.1-5.fc25', <br>   url => $magnum_proto//localhost:8888 <br> } | $magnum_proto = ''http:'' does not reach url |

variable $magnum_proto. This string expression is an instance of HTTP without TLS. As shown in the 'Reachability' column, $magnum_proto reaches the url attribute, because between $magnum_proto and url there exists no code element that changes the value of $magnum_proto. On the other hand, as shown in line#1,#2 of the second row, $magnum_proto does not reach url as the value of $magnum_proto is changed from 'http' to 'ftp:'.

Each DDG has three types of nodes: taint, intermediate, and sink. Taint nodes correspond to variables with security weaknesses that are identified in Section 2.1.2. Sink nodes correspond to attributes used within a resource. Intermediate nodes are non-taint and non-sink nodes that are reachable from one or multiple taint nodes. By construction a DDG will include at least one taint node and at least one sink node. A DDG includes $>= 0$ intermediate nodes.

## 2.2 Methodology for RQ$_1$

We answer RQ$_1$ by computing the detection accuracy of TaintPup with metrics, such as precision [62]. An increase in precision for TaintPup compared to that of SLIC will provide evidence on how detection propagation aids in security weakness identification. We use the following steps in this regard:

### 2.2.1 Dataset Construction

We use OSS repositories mined from GitHub, GitLab, and three IT organizations who use Puppet manifests to manage their computing infrastructure: Mozilla, Openstack, and Wikimedia Commons.

OSS repositories are susceptible to quality issues, which necessitates curation of the collected repositories [44]. In prior work [6], [34], [44], researchers have leveraged a set of attributes to filter OSS GitHub repositories reflective of professional software development. These attributes include count of Puppet manifests [52], count of commits per month [44], and count of contributors [6], [34]. Taking motivation from prior work, we apply the following filtering criteria: *Criterion-1*: The proportion of Puppet manifests is $>= 10\%$. IaC scripts can co-locate with other types of files, such as source code files and build files [31]. We assume that with this threshold we can exclude repositories that do not have sufficient Puppet manifests for analysis. *Criterion-2*: The repository is not a copy of another to avoid duplicates. *Criterion-3*: Count of contributors is $>= 10$. Similar to prior work [53], [54], we use this criterion to filter repositories used for personal purposes, such as coursework. *Criterion-4*: Lifetime of the repository is $>= 1$ month. Using this criterion, we filter repositories with short lifetime. We measure lifetime by calculating the difference between the last commit date and the creation date for the repository. *Criterion-5*: The repository has $>= 25$ commits to filter repositories with limited activity. *Criterion-6*: The repository has $>= 2$ commits per month. Munaiah et al. [44] used this threshold to identify mature OSS GitHub repositories.

We collect the Mozilla, Openstack, and Wikimedia repositories from their corresponding public repository databases (Mozilla [43], Openstack [46], Wikimedia [64]). We use Google BigQuery [27] to download OSS repositories hosted on GitHub that use Puppet. We use the GitLab API [22] to mine OSS repositories hosted on GitLab. Table 4 summarizes how many repositories are filtered using our criteria. We download 336 repositories by cloning the master branches on October 2021. Attributes of collected repositories is available in Table 5.

TABLE 4: Repository Filtering

|  | GitHub | GitLab | Mozilla | Openstack | Wiki. |
|---|---|---|---|---|---|
|  | 3,405,303 | 1,659 | 1,594 | 2,262 | 2,509 |
| Criterion-1 | 18,187 | 38 | 2 | 96 | 13 |
| Criterion-2 | 17,872 | 38 | 2 | 96 | 13 |
| Criterion-3 | 856 | 30 | 2 | 94 | 13 |
| Criterion-4 | 770 | 30 | 2 | 90 | 11 |
| Criterion-5 | 675 | 30 | 2 | 90 | 11 |
| Criterion-6 | 241 | 25 | 2 | 61 | 7 |
| **Final** | 241 | 25 | 2 | 61 | 7 |

TABLE 5: Dataset Attributes

| Attribute | GitHub | GitLab | Mozilla | Openstack | Wiki. |
|---|---|---|---|---|---|
| Total Repos. | 241 | 25 | 2 | 61 | 7 |
| Total Commits | 599,900 | 1,943 | 2 | 42,446 | 16,231 |
| Average Duration (Month) | 241 | 34.2 | 90 | 38.5 | 60.0 |
| Total Puppet Manifests | 11,477 | 885 | 1,613 | 2,952 | 704 |
| Total Puppet LOC | 498,241 | 49,430 | 66,367 | 234,640 | 27,889 |
| Total Distinct Resources | 65,599 | 5,055 | 10,583 | 23,754 | 3,561 |

### 2.2.2 Dataset Labeling

We use three steps to perform labeling:

*Step#1-Rater Training*: A software engineer, who is not an author of the paper, volunteered to participate in labeling all 17,629 Puppet manifests. The rater has an experience of one year in software engineering and cybersecurity. Before applying labeling, we conduct a training session where the rater is mentored by the first author. The training session was conducted in two phases: *first in phase-1.1*, both the first author and the rater independently inspect a randomly-selected set of 500 Puppet manifests. The first author and the rater use a (i) guidebook [51] with names, definitions, and examples of security weakness categories, and (ii) the online documentation of Puppet [35] that describes syntax and information flow in Puppet manifests. The rater and the first author individually determines if a security weakness is in fact true positive by first inspecting if a security weakness exists, and then if the weakness is used by $>= 1$ resources. Upon completion of the inspection process the rater discuss their agreements and disagreements. At this stage the Cohen's Kappa is 0.47, indicating 'moderate' agreement according to Landis and Koch [36]. The disagreement occurred due to the rater's misunderstanding of a weakness being used by a resource. *Second in phase-1.2*, upon discussion of their disagreements, the rater and the first author conducted another round of inspection. The process and used materials are similar to that of phase-1.1. At this stage, Cohen's Kappa is 1.0 between the rater and the first author, which gives us the confidence that the rater is equipped with necessary background to label all 17,629 manifests.

*Step#2-Labeling*: Similar to the training session, the rater uses the guidebook and the Puppet online documentation [35] to identify security weaknesses in 17,629 Puppet manifests. One manifest can include multiple categories of security weaknesses, and thus the rater can map one Puppet manifest to one or multiple of the six categories. The rater takes 745 hours to complete labeling. Altogether, the rater identifies 4,906 security weaknesses. Count of security weaknesses for the five datasets is presented in Table 6.

TABLE 6: Count of Security Weaknesses in Our Datasets

| Category | GitHub | GitLab | Mozilla | Openstack | Wiki. |
|---|---|---|---|---|---|
| Admin by default | 5 | 0 | 0 | 12 | 0 |
| Empty password | 40 | 0 | 2 | 3 | 21 |
| Hard-coded secret | 2,604 | 105 | 145 | 751 | 63 |
| Invalid IP address binding | 31 | 1 | 12 | 62 | 0 |
| Use of HTTP without TLS | 543 | 7 | 5 | 465 | 22 |
| Use of weak crypto. algo. | 3 | 2 | 0 | 2 | 0 |
| **Total** | 3,226 | 115 | 164 | 1,295 | 106 |

*Step#3-Rater Verification*: We use a PhD student in the department, who is not an author of the paper, to verify the rater's labeling. We use a randomly-selected set 500 manifests from our set of 17,629 manifests that is not used in Step#1. Similar to Step#1, the PhD student is provided the guidebook and Puppet's online documentation [35]. Upon completion, we record a Cohen's Kappa of 0.91 between the PhD student and the rater.

### 2.2.3 Evaluate TaintPup's Detection Performance

We evaluate TaintPup's detection performance by applying the following steps: *first*, we run TaintPup and SLIC on the collected 17,629 manifests. *Second*, we use three metrics: precision, recall, and F-measure, similar to prior work [47]. Precision refers to the fraction of correctly identified instances among the total identified security weaknesses, as determined by a static analysis tool. Recall refers to the fraction of correctly identified instances retrieved by a static analysis tool over the total amount instances. F-measure is the harmonic mean of precision and recall [62].

## 2.3 Answer to RQ$_1$: TaintPup's Detection Accuracy

We answer **RQ$_1$: How does propagation detection improve security weakness identification in Puppet manifests?** in this section. We report the precision and F-measure for SLIC and TaintPup with Tables 7 and 8. With respect to precision, TaintPup outperforms SLIC for all categories across all datasets. According to Table 7, TaintPup's average precision is 3.3, 2.5, 2.4, 3.4, and 1.5 times higher than that of SLIC respectively, for GitHub, GitLab, Mozilla, Openstack, and Wikimedia. Considering 4,906 security weaknesses across all five datasets, the average precision is 2.4 times higher than that of SLIC. Furthermore, across all five datasets the average F-measure is 1.8 times higher than that of SLIC. We observe a recall of 1.0 for both SLIC and TaintPup for all six categories. This shows TaintPup's ability to detect all identified security weaknesses with higher precision than that of SLIC, without reducing recall.

## 3 METHODOLOGY FOR EMPIRICAL STUDY

Methodology to answer RQ$_2$, RQ$_3$, and RQ$_4$ is described below.

TABLE 7: Answer to RQ$_1$: Precision of SLIC and TaintPup for Six Security Weakness Categories

| | GitHub | | GitLab | | Mozilla | | Openstack | | Wikimedia | |
|---|---|---|---|---|---|---|---|---|---|---|
| Category | SLIC | TaintPup | SLIC | TaintPup | SLIC | TaintPup | SLIC | TaintPup | SLIC | TaintPup |
| Admin by default | 0.15 | 0.83 | NA | NA | NA | NA | 0.10 | 0.86 | NA | NA |
| Empty password | 0.06 | 0.93 | NA | NA | 0.30 | 1.00 | 0.08 | 0.75 | 0.75 | 0.84 |
| Hard-coded secret | 0.63 | 0.95 | 0.55 | 0.92 | 0.50 | 0.96 | 0.49 | 0.94 | 0.44 | 0.83 |
| Invalid IP address binding | 0.20 | 0.89 | 0.25 | 1.00 | 0.62 | 0.86 | 0.26 | 0.93 | NA | NA |
| Use of HTTP without TLS | 0.53 | 0.97 | 0.47 | 0.88 | 0.11 | 0.83 | 0.59 | 0.96 | 0.52 | 0.88 |
| Use of weak crypto. algo. | 0.03 | 0.60 | 0.27 | 1.00 | NA | NA | 0.07 | 1.00 | NA | NA |
| **Average** | 0.26 | 0.86 | 0.38 | 0.95 | 0.38 | 0.91 | 0.26 | 0.90 | 0.57 | 0.85 |

TABLE 8: Answer to RQ$_1$: F-measure of SLIC and TaintPup for Six Security Weakness Categories

| | GitHub | | GitLab | | Mozilla | | Openstack | | Wikimedia | |
|---|---|---|---|---|---|---|---|---|---|---|
| Category | SLIC | TaintPup | SLIC | TaintPup | SLIC | TaintPup | SLIC | TaintPup | SLIC | TaintPup |
| Admin by default | 0.27 | 0.91 | NA | NA | NA | NA | 0.18 | 0.92 | NA | NA |
| Empty password | 0.12 | 0.93 | NA | NA | 0.46 | 1.00 | 0.15 | 0.86 | 0.86 | 0.91 |
| Hard-coded secret | 0.77 | 0.97 | 0.81 | 0.96 | 0.67 | 0.98 | 0.66 | 0.97 | 0.69 | 0.79 |
| Invalid IP address binding | 0.33 | 0.94 | 0.40 | 1.00 | 0.76 | 0.92 | 0.42 | 0.96 | NA | NA |
| Use of HTTP without TLS | 0.69 | 0.98 | 0.62 | 0.93 | 0.21 | 0.91 | 0.75 | 0.98 | 0.68 | 0.94 |
| Use of weak crypto. algo. | 0.05 | 0.75 | 0.54 | 1.00 | NA | NA | 0.07 | 1.00 | NA | NA |
| **Average** | 0.37 | 0.91 | 0.59 | 0.97 | 0.52 | 0.95 | 0.37 | 0.95 | 0.74 | 0.88 |

### 3.1 Methodology to Answer RQ$_2$

We use RQ$_2$ to characterize how frequently security weaknesses propagate into resources. Such characterization can help us understand how many of the security weaknesses are actually being used by a resource. We answer RQ$_2$ by *first* reporting the total count of resources in each dataset into which security weaknesses propagate. *Second*, with Equation 1 we compute 'Impacted Resource (%)', i.e., the proportion of resources in each dataset for which $>= 1$ security weakness propagates. We use 'Impacted Resource', as resource is the fundamental unit to specify configurations in order to manage a computing infrastructure [35]. If we can demonstrate empirical evidence that the identified security weaknesses are actually used by resources, then practitioners will be informed on how security weaknesses are used, and the resources they are used in. *Third*, we report the minimum, median, and maximum number of resources in a manifest into which a security weakness propagate. To answer RQ$_2$, we use TaintPup as it allows us to identify security weaknesses used by attributes within resources with the help of DDGs.

$$\text{Impacted Resource}(\%) = \frac{\text{\# of resources in which} >= 1 \text{ security weakness propagates}}{\text{total \# of unique resources in the dataset}} * 100\% \quad (1)$$

### 3.2 Methodology to Answer RQ$_3$

We conduct a categorization of affected resources to gain further understanding of the resources affected by security weaknesses. We apply the following steps: *first*, we identify affected attributes and resources from the output of TaintPup. *Second*, we identify the titles and types from the affected resources. *Third*, we apply open coding [57] with titles and types of affected resources. Open coding is a qualitative analysis technique that is used to identify categories from structured or unstructured text [57]. *Fourth*, we compute the proportion of resources that belong to a certain category.

*Rater Verification*: The first author who has 6 years of experience in Puppet development performs open coding. The derived categories are susceptible to rater bias, which we mitigate by using a PhD student in the department, who is not an author of the paper. The additional rater was assigned 100 randomly-selected resources for mapping them to the categories identified by the first author. We record a Cohen's Kappa [15] of 0.87 between the PhD student and the first author, indicating 'perfect' agreement [36].

### 3.3 Methodology to Answer RQ$_4$

We answer RQ$_4$ by conducting an online survey with practitioners who have developed Puppet manifests. We contact these practitioners via e-mails, which we mine from the 336 OSS repositories reported in Section 2.2.1. We randomly select 250 e-mail addresses, which we use to send the surveys. We offer a drawing of two 50 USD Amazon gift cards as an incentive for participation following Smith et al. [60]'s recommendations. We conduct the survey from December 2021 to March 2022 following the Internal Review Board (IRB) protocol #2356.

In the survey, we first ask developers about their experience in developing Puppet manifests. Next, we describe each of the identified resource categories with definitions and examples. We then ask questions related to perceived frequency and severity: *first*, we ask "*How frequently do you think the identified resource categories are affected by security weaknesses?*" Survey participants used a five-item Likert scale to answer this question: 'Not at all frequent', 'Rarely', 'Somewhat frequently', 'Frequently', and 'Highly frequent'.

*Second*, we ask *"What is the severity of the identified resource categories into which security weaknesses appear?"* To answer this question, survey participants used the following five-item Likert scale: 'Not at all severe', 'Low severity', 'Moderately severe', 'Severe', and 'Highly severe'. We use a five-item Likert scale for both questions following Kitchenham and Pfleeger's guidelines [33]. Furthermore, following Kitchenham and Pfleeger [33]'s advice we apply the following actions before deploying the survey: (i) provide an estimate of completion time, (ii) provide explanations related to the purpose of the study, (iii) provide survey completion instructions, and (iv) provide explanations confirming preservation of confidentiality. The survey questionnaire is included in our verifiability package [51].

## 4 EMPIRICAL FINDINGS

We provide answers to $RQ_2$, $RQ_3$, and $RQ_4$ as follows:

### 4.1 Answer to $RQ_2$: Propagation Frequency

In this section, we provide answer to **$RQ_2$: How frequently do security weaknesses propagate into resources?** Altogether, we identify 4,906 security weaknesses to propagate into 4,457 distinct resources. The count of resources into which weaknesses propagate is: 2,945 for GitHub, 104 for GitLab, 167 for Mozilla, 1,128 for Openstack, and 113 for Wikimedia. Considering all datasets, security weaknesses propagate into 4.1% of 108,552 resources. The 'Impacted Resource (%)' column shows the proportion of resources into which $>= 1$ security weaknesses propagate. For example, for GitHub, we observe 3,872 security weaknesses to propagate into 4.49% of 65,599 resources. The proportion of affected resources is highest for Openstack. Details are available in Table 9.

TABLE 9: Answer to $RQ_2$ : Frequency of Resources Into Which Security Weaknesses Propagate

| | Impacted Resource (%) | | | | |
|---|---|---|---|---|---|
| **Category** | **GitH.** | **GitL.** | **Mozilla** | **Open.** | **Wiki.** |
| Admin by default | 0.01 | NA | 0.00 | 0.05 | NA |
| Empty password | 0.09 | 0.02 | 0.02 | 0.02 | 0.05 |
| Hard-coded secret | 3.55 | 1.82 | 1.30 | 3.25 | 2.22 |
| Invalid IP address binding | 0.05 | 0.02 | 0.17 | 0.19 | 0.00 |
| Use of HTTP without TLS | 0.79 | 0.16 | 0.09 | 1.24 | 0.89 |
| Use of weak crypto. algo. | 0.01 | 0.02 | NA | 0.004 | NA |
| **Combined** | **4.49** | **2.06** | **1.58** | **4.75** | **3.17** |

In Table 10, we report the minimum, median, and maximum number of resources in a manifest into which $>= 1$ security weakness propagates. We observe a security weakness can propagate into as many as 35 distinct resources.

### 4.2 Answer to $RQ_3$: Resource Categories

We provide answers to **$RQ_3$: What are the resource categories into which security weaknesses propagate?** by first describing the resource categories in Section 4.2.1. Next, we report frequency of resource categories in Section 4.2.2.

TABLE 10: Answer to $RQ_2$: Resource Frequency (Minimum, Median, Maximum)

| Category | GitHub | GitLab | Mozilla | Openstack | Wiki. |
|---|---|---|---|---|---|
| Admin by default | 1, 1, 2 | NA | 0, 0, 0 | 1, 1, 1 | NA |
| Empty password | 1, 2, 5 | 1, 1, 1 | 1, 1, 1 | 1, 1, 1 | 1, 1, 1 |
| Hard-coded secret | 1, 1, 35 | 1, 1, 6 | 1, 1, 6 | 1, 1, 12 | 1, 1, 4 |
| Invalid IP address binding | 1, 1, 2 | 1, 1, 1 | 1, 3, 5 | 1, 1, 3 | 0, 0, 0 |
| Use of HTTP without TLS | 1, 1, 7 | 1, 3, 4 | 1, 2.5, 4 | 1, 1, 6 | 1, 1, 6 |
| Use of weak crypto. algo. | 1, 1, 1 | 1, 1, 1 | NA | 1, 1, 1 | NA |
| **Total** | 1, 1, 35 | 1, 1, 6 | 1, 1, 6 | 1, 1, 12 | 1, 1, 6 |

#### 4.2.1 Description of Resource Categories

We identify 7 categories of resources into which security weaknesses propagate. We describe each category with examples as follows.

**I-*Communication Platforms***: Resources used to manage communication platforms, such as Discourse [1] and Slack [2].

*Example*: Listing 1 shows how a hard-coded user name is used to manage Slack contacts in line# 9. The hard-coded username is $slack_username = 'Icinga', which is later used by the resource icinga::slack_contact.

```
1 $notify_slack = false,
2 $notify_graphite = true,
3 $slack_channel = undef,
4 $slack_username = 'Icinga',
5 ---
6 icinga::slack_contact { 'slack_search_team':
7     slack_webhook_url       =>
        $slack_webhook_url,
8     slack_channel           =>
        '#govuk-searchandnav',
9     slack_username          =>
        $slack_username,
10    icinga_status_cgi_url   =>
        $slack_icinga_status_cgi_url,
11    icinga_extinfo_cgi_url  =>
        $slack_icinga_extinfo_cgi_url,
12 }
```

Listing 1: A hard-coded secret propagating into a resource used to manage Slack.

**II-*Containerization***: Resources used to manage containers.

*Example*: Listing 2 shows an example of a resource that is used to perform authentication for the Magnum container service [3]. We observe an instance of insecure HTTP for $magnum_protocol, which is later used for $magnum_url in line#3. $magnum_protocol is also used by two attributes in the magnum resource, but $magnum_url is not used by any resource within the manifest. TaintPup is able to accurately detect both: the propagation of $magnum_protocol into the magnum resource, as well as $magnum_url not propagating at all.

**III-*Continuous Integration***: Resources used to manage infrastructure needed to implement the practice of continu-

1. https://www.discourse.org/
2. https://slack.com/
3. https://wiki.openstack.org/wiki/Magnum

```
1 $magnum_protocol = 'http'
2 ...
3 $magnum_url =
  ↪ "${magnum_protocol}://${magnum_host}:$magnum_port/v1"
4 magnum { '::magnum::keystone::authtoken':
5   auth_uri  =>
    ↪ "${magnum_protocol}://${magnum_host}:5000/v3",
6   auth_url  =>
    ↪ "${magnum_protocol}://${magnum_host}:35357",
7   ...
8 }
```

Listing 2: An instance of insecure HTTP propagating into a resource used to manage Magnum-based containers.

ous integration (CI), with tools, such as Jenkins [30]. CI tools integrate code changes by automatically compiling, building, and executing test cases upon submission of code changes [18].

*Example*: In Listing 3 an instance of empty password propagates into a resource to setup configurations for Jenkins. As shown in line #9, the empty password instance $jenkins_management_password = '' is used to construct $security_opt_params using join, a Puppet function used to concatenate strings [35]. Later with the exec resource, $security_opt_params is used to manage configurations for a Jenkins-based CI infrastructure.

```
1  class jenkins::master (
2    ...
3    $jenkins_management_password  = '',
4    ...
5    $security_opt_params = join([
6      'set_security_password',
7      "'${jenkins_management_login}'",
8      "'${jenkins_management_email}'",
9      "'${jenkins_management_password}'",
10     "'${jenkins_management_name}'",
11     "'${jenkins_ssh_public_key_contents}'",
12     "'${jenkins_s2m_acl}'",
13   ], ' ')
14   ...
15   exec { 'jenkins_auth_config':
16     require  => [
17       File["${jenkins_libdir}/jenkins_cli.groovy"],
18       Package['groovy'],
19       Service['jenkins'],
20     ],
21     command  => join([
22       '/usr/bin/java',
23       "-jar ${jenkins_cli_file} -s",
        ↪ "${jenkins_proto}://${jenkins_address}:"
24       "${jenkins_port}",
25       "groovy
        ↪ ${jenkins_libdir}/jenkins_cli.groovy",
26       $security_opt_params,
27     ], ' '),
28     tries    => $jenkins_cli_tries,
29     ...
30   }
```

Listing 3: An empty password propagating into a resource used to manage Jenkins.

IV-***Data Storage***: Resources used to manage data storage systems, such as MySQL servers, PostgreSQL servers, and Memcached.

*Example*: Listing 4 shows an instance of empty password used by a resource to manage a MySQL database. The mysql::db resource uses $database_password='' with the password attribute.

```
1  class gerrit::mysql(
2    ...
3    $database_password = '',
4  ) {
5    mysql::db { $database_name:
6      ...
7      password => $database_password,
8      host     => 'localhost',
9      grant    => ['all'],
10     ...
11   }
12   ...
13 }
```

Listing 4: An instance of empty password propagating into a resource used to manage a MySQL database.

V-***File***: Resources used to manage files by performing file-related operations, such as reading, writing, or deleting a file. We observe security weaknesses to propagate in Puppet-defined resources, such as file, and custom resources.

*Example*: Listing 5 shows how SHA1 is used to encrypt a password with the htpasswd_sha1 function. The encrypted password is assigned to $nagiosadmin_pw, which is later used to manage a file with the File['nagios_htpasswd'] resource, as shown in line#5.

```
1 $nagiosadmin_pw  =
  ↪ htpasswd_sha1($nagios_hiera['nagiosadmin_pw'])
2 $nagios_hosts    = $nagios_hiera['hosts']
3 File['nagios_htpasswd'] {
4   source  => undef,
5   content => "nagiosadmin:${nagiosadmin_pw}",
6   mode    => '0640',
7 }
```

Listing 5: An instance of SHA1 usage propagating into a resource used to manage a file.

VI-***Load Balancers***: Resources used to manage load balancers, such as HAProxy [24]. Load balancing is used to systematically distribute network or application traffic across multiple servers [12].

*Example*: Listing 6 shows an example of a security weakness propagating into a resource used to manage HAProxy. $vip is an instance of an invalid IP address, which is used by $api_vip_orig and $discovery_vip_orig respectively, in lines #8 and 14. Both $api_vip_orig and $discovery_vip_orig will be assigned '0.0.0.0' with $vip through the execution of the else block as both $api_server_vip and $discovery_server_vip is assigned undef, which is false when used as Boolean. $api_vip_orig and $discovery_vip_orig are respectively, used in lines #17 and #21 to manage HAProxy services. Listing 6 is an example that illustrates TaintPup's ability to detect the propagation of one security weakness into multiple resources.

VII- ***Networking***: Resources used to manage network-related functionalities, such as setting up firewalls, network controllers, and managing virtual local area networks.

*Example*: Listing 7 shows an example of a hard-coded password propagating into a resource used for management of network infrastructure. The resource is used to

```
1 $vip                           = '0.0.0.0',
2 $api_server_vip                = undef,
3 $discovery_server_vip          = undef,
4 ...
5 if $api_server_vip {
6    $api_vip_orig = $api_server_vip
7 } else {
8    $api_vip_orig = $vip
9 }
10
11 if $discovery_server_vip {
12    $discovery_vip_orig = $discovery_server_vip
13 } else {
14    $discovery_vip_orig = $vip
15 }
16 rjil::haproxy_service { 'api':
17    vip                         => $api_vip_orig,
18    ...
19 }
20 rjil::haproxy_service { 'discovery':
21    vip                         => $discovery_vip_orig,
22    ...
23 }
```

Listing 6: An instance of invalid IP address propagating into a resource used to manage HAProxy, a load balancer.

manage the Open Network Operating System (ONOS) controller [45]. The hard-coded password $password = 'karaf' is used by $dashboard_desc, which is later used to construct $json_hash. In line #10, $json_hash is used by $json_messsage. Later, as shown in line#12, the exec resource uses $json_message to execute a command in line#12 in order to create an ONOS dashboard link.

```
1 notice(' ONOS MODULAR: onos-dashboard.pp')
2 ...
3 $password = 'karaf'
4 ...
5 $dashboard_desc = "Onos dashboard interface.
   ↪ Default credentials are ${user}/${password}"
6
7 $json_hash = { title        => $dashboard_name,
8                description  => $dashboard_desc,
9                url          => $dashboard_link, }
10 $json_message = $json_hash
11 exec { 'create_dashboard_link':
12    command => "/usr/bin/curl -H 'Content-Type:
      ↪ application/json' -X POST -d
      ↪ '${json_message}'
      ↪ https://${master_ip}:8000/api/clusters/${cluster_id}"
13 }
```

Listing 7: A hard-coded password propagating into a resource used to manage ONOS.

### 4.2.2 Frequency of Affected Resource Categories

Findings from Table 11 show that security weaknesses are in fact used for managing infrastructure, such as CI, container, and data storage infrastructure. For example, for GitHub 69% of identified hard-coded secrets are used to manage CI-based infrastructure. A complete breakdown is available in Table 11, where we provide a mapping between each security weakness and resource categories into which security weaknesses propagate. 'CI', 'Comm', 'Container', 'Data', 'File', 'Load', and 'Network' respectively refers to continuous integration, communication platforms, containerization, data storage, file, load balancer, and network. A resource category name is followed by the proportion of security weaknesses that propagate into resources for that category. 'NA' indicates a security weakness to not propagate into a resource for a certain category. The percentage of affected resource categories is listed in Figure 3. For example, the total count of affected resources by security weaknesses is 2,945 for the Github dataset, of which 0.1% are used to manage communication platforms. Our findings provide a cautionary tale on the state of Puppet manifest security, as we observe identified security weaknesses to propagate into resources for infrastructure management, which in turn leaves computing infrastructure susceptible to security attacks.

| Reso. Categ. | GitHub | GitLab | Mozi. | Ostk. | Wiki. |
|---|---|---|---|---|---|
| Network | 8.5 | 3.1 | 0 | 6 | 11.4 |
| Load Balance | 0.5 | 2.4 | 0 | 29.3 | 11.1 |
| File | 9.1 | 47.7 | 61.7 | 22.6 | 33.7 |
| Data Storage | 9.3 | 35.1 | 6.6 | 19.1 | 19.5 |
| Container | 3.8 | 0 | 24.4 | 17.7 | 11.3 |
| Comm. Platform | 0.1 | 0.6 | 0 | 0.9 | 0 |
| CI | 68.7 | 11.1 | 7.3 | 4.4 | 13 |

Fig. 3: Answer to RQ$_3$: Percentage of Affected Resources.

### 4.3 Answer to RQ$_4$: Practitioner Perception

In this section, we provide answer to **RQ$_4$: What are the practitioner perceptions of the identified resources into which security weaknesses propagate?**. From our survey, we obtain 24 responses in total, Of the 24 practitioners, 11, 1, 1, 2, and 9 practitioners respectively had an experience of < 1, 1 − 2, 3 − 4, 4 − 5, and > 5 years of experience in Puppet. In Figures 4 and 5 we respectively, report practitioner perceptions for frequency and severity of the identified resource categories. The x and y-axis respectively presents the percentage of survey participants and resource categories. For example, from Figure 4 we observe 25% of the total survey respondents to identify containerization as a resource category for which security weaknesses frequently or highly frequently propagate.

From Figure 4, we observe survey respondents to perceive CI management to be most frequently affected by security weaknesses. Such perception is congruent with the GitHub-related findings presented in Figure 3, where we observe resources related to CI to be the most frequent category. Furthermore, based on Figure 3 we observe the proportion of resources related to management of communication platforms to be < 1.0%. From Figure 5, we observe propagation of security weaknesses for data storage management to be perceived as most severe.

TABLE 11: Answer to RQ$_3$: Mapping of Resource Categories and Security Weaknesses

| Category | GitHub | GitLab | Mozilla | Openstack | Wikimedia |
|---|---|---|---|---|---|
| Admin by default | Data:100% | NA | NA | Container:11.1%, Data:27.8%, File:5.5%, Load:27.8%, Network:27.8% | NA |
| Empty password | CI:62.0%, Data:32.4%, File:5.6% | File:100% | File:100% | Data:40.0%, Load:60.0% | Data:100% |
| Hard-coded secret | CI:69%, Comm:0.1%, File:9.3%, Network:8.9% | Container:3.6%, Data:8.8%, Load:0.3% | File:45%, Data:55% | CI:10.0%, Container:21.9%, Data:0.3%, File:67.8% | CI:3.5%, Container:18.7%, Comm:1.1%, Data:38.4%, File:31.6%, Load:2.6%, Network:4.1% | CI:11.1%, Container:1.6%, Data:78.6%, File:5.5%, Load:0.8%, Network:2.4% |
| Invalid IP address binding | CI:2.4%, Data:43.9%, Load:21.9%, Network:22.1% | Container:7.3%, File:2.4% | Data:100% | Network:100% | Container:18.6%, Data:40.6%, File:6.8%, Load:28.9%, Network:5.1% | NA |
| Use of HTTP without TLS | CI:49.6%, Data:16.5%, Load:5.7%, Network:2.1% | Container:22.2%, File:3.9% | Container:100% | Container:100% | CI:7.1%, Container:15.6%, Comm:0.7%, Data:31.1%, File:4.8%, Load:31.0%, Network:9.7% | CI:2.1%, Data:85.4%, File:12.5% |
| Use of weak crypto. algo. | File:50%, Data:50% | File:100% | NA | Data:100% | NA |

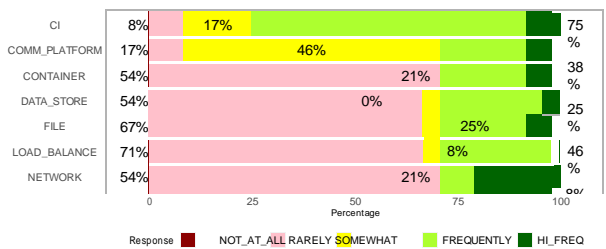

Fig. 4: Answer to RQ$_4$: Practitioner perception of frequency for identified resource categories.

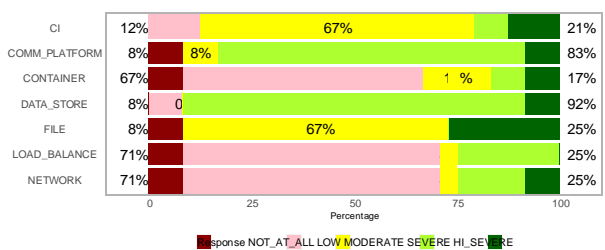

Fig. 5: Answer to RQ$_4$: Practitioner perception of severity for identified resource categories.

Figures 4 and 5 show nuanced perspectives from practitioners. For example, while 88% of respondents perceive CI-related resources to be frequently impacted by security weaknesses, 12% of total respondents find security weaknesses in CI-based management to have severe or highly severe impact. Also, 92% of the respondents identify security weakness propagation for data storage-related resources to be severe or highly severe, but only 46% of the respondents perceive such propagation to be frequent.

## 5 DISCUSSION

We discuss the implications of our paper in this section.

***Implications Related to Practitioner Actionability:*** Our results in Table 7 show TaintPup to have higher precision than that of SLIC, which is a state-of-the-art security static analysis tool for Puppet. Low precision static analysis tool contribute to a lack of actionability, which in turn results in abandonment of static analysis tool usage [10], [26], [32]. Unlike SLIC, TaintPup generates fewer false positives. On average, TaintPup's precision is 2.4 times higher than that of SLIC. We attribute TaintPup's precision improvement to the advanced syntax analysis, and application of information flow analysis. As better precision is correlated with increased actionability for static analysis tools [25], [56], TaintPup's precision improvement can help practitioners take actions to mitigate security weaknesses.

Another utility of TaintPup is its capability to report the flow of a detected security weakness, and whether or not it is being used by a resource. Unlike SLIC, TaintPup reports the name and location of a manifest, the resources affected by a security weakness, and the attribute used by the resource into which the security weakness propagate. Such capability gives practitioners the ability to assess if a detected security weakness is relevant or not. We recommend use of information flow analysis to detect security weaknesses in Puppet manifests because it (i) identifies resources that are affected by security weaknesses, and (ii) reduces false positives.

***Survey-related Implications:*** Implications of RQ$_4$ are: **Severity-related Perceptions:** Findings reported in Section 4.3 show practitioners to not identify security weakness propagation for all resource categories to be severe. According to Figure 5, practitioners perceive propagation of security weaknesses to be least severe for CI and container infrastructure management. However, these perceptions could leave unmitigated security weaknesses during management of CI and containers, which in turn could be used by malicious users to perform cryptomining attacks [38]. Existence of security weaknesses in CI infrastructure resulted in the Codecov incident, which impacted 29,000 customers, and breached hundreds of customer networks [1], [2], [4]. According to Table 11, security weaknesses, such as hard-coded secrets propagate into container-related resources, which in turn can cause container escape, where a container user is able to nullify container isolation and access unauthorized resources [40]. Container escapes motivated malicious users conduct security attacks on container-based infrastructure, as many as 17,358 attacks in 18 months [3], [5].

**Frequency-related Perceptions:** Our frequency-related findings in Section 4.3 show a disconnect between what practitioners perceive, and empirical results. While practitioners perceive file-related resources to be least frequently affected by security weaknesses according to Figure 4, these resources are most frequently affected for GitLab, Mozilla, and Wikimedia as shown in Figure 3. These findings suggest a lack of practitioner awareness on how frequently security weaknesses affect Puppet-based infrastructure management, which can be mitigated through the use of TaintPup.

***Implications Related to Prioritizing Inspection Efforts:*** Our empirical study has implications for prioritizing inspection efforts as well. While conducting security focused code reviews, practitioners can focus on the resources for which TaintPup reports a security weakness. In this manner, instead of inspecting all resources, with the help of TaintPup practitioners can inspect a smaller set of resources.

***Future Work*** We discuss opportunities for future work:
**Improvement Opportunities for TaintPup**: TaintPup can be extended so that practitioners themselves can specify the sinks to track security weaknesses in Puppet manifests. Currently, TaintPup uses attributes in resources as sinks, which could be limiting because a security weakness can be used by a code snippet deemed important by practitioners, but is not an attribute.

**Security-focused Information Flow Analysis for Other IaC Languages**: Security-focused information flow analysis for other languages is an opportunity for future work. Such analysis will require an understanding of what code elements are used to manage infrastructure in other languages. For example, with respect to syntax and semantics Puppet is different from Ansible [54], which requires information flow analysis tools tailored for Ansible manifests.

## 6 THREATS TO VALIDITY

**Conclusion Validity**: TaintPup builds DDGs leveraging def-use chains [7], which may not capture all types of information flow. For example, if a hard-coded password is provided as a command line input or as a catalog [35], then TaintPup will not report a security weakness. Security weakness categories determined by TaintPup are limited to Rahman et al. [53]'s paper.

**Construct Validity**: In Section 2.2.2, when determining security weaknesses the rater may have implicit biases that could have affected the labeling for the five datasets. We mitigate this limitation by allocating a rater who is not the author of the paper, and also by performing rater verification.

**External Validity**: Our empirical study is susceptible to external validity as our analysis is limited to datasets collected from OSS repositories. TaintPup can generate false positives and false negatives for datasets not used in the paper, which in turn can influence results presented in Sections 4.1 and 4.2.

## 7 RELATED WORK

Our paper is related to prior research on Puppet-related code elements that are indicative of quality concerns. Sharma et al. [59], Bent et al. [63], and Rahman and Williams [8] in separate studies identified Puppet-related code elements that are indicative of defects in Puppet manifest. Analysis of specific defects, such as security defects has garnered interest amongst researchers too. By mining OSS repositories Rahman et al. [52] found absence of Puppet code elements to cause security defects. Existence of security defects, such as security weaknesses were further confirmed by Rahman et al. [53], where they identified seven categories of security weaknesses. They further replicated the study in another paper [54], where they observed security weaknesses in Puppet manifests to also appear for Ansible manifests. Rahman et al. [53]'s paper was also replicated by Hortlund [28], who reported the security weakness density to be less than that of reported by Rahman et al., due to false positives generated by SLIC. Bhuiyan and Rahman [11] reported similar observations: they manually inspected 2,764 Puppet manifests, and documented SLIC to generated 1,560 false positives.

Our paper is also related to prior research that has applied taint tracking for quality analysis of Android, Java, and Python applications. Xia et al. [65] used taint tracking to build AppAudit. Using AppAudit, they [65] found most data leaks to be caused by third-party advertising modules. Gibler et al. [21] performed taint tracking to identify 57,299 privacy leaks in 7,414 Android apps. Arzt et al. [9] applied alias-based taint tracking to construct FlowDroid so that leaks are detected in 500 Android apps. Mahmud et al. [39] used taint tracking to identify Android apps that violate Payment Card Industry (PCI) compliance standards. Gordon et al. [23] applied taint analysis to detect inter component communication (ICC) leaks in 24 Android apps. Java-specific taint tracking tools have also been proposed. Chin and Wagner [14] conducted character level taint tracking to detect vulnerabilities in Java web-based applications. Conti and Russo [16] constructed a Python-based taint tracking tool to identify vulnerabilities in Python applications. Peng et al. [48] used taint tracking to verify integrity of Python applications. However, none of these tools are applicable for Puppet manifests as they do not account for Puppet's state-based infrastructure management approach as well as code elements unique to Puppet.

From the above-mentioned discussion we observe propagation of security weaknesses in Puppet manifests to be an under-explored research area, as none of the above-mentioned papers investigate how security weaknesses impact Puppet-based infrastructure management. We address this research gap by constructing TaintPup, and then we use TaintPup to conduct an empirical study.

## 8 CONCLUSION

While IaC scripts, such as Puppet manifests have yielded benefits for managing computing infrastructure at scale, these manifests include security weaknesses, such as hard-coded passwords and use of weak cryptography algorithms. To detect and characterize security weaknesses propagation for Puppet-based infrastructure management, we have constructed TaintPup, using which we conduct an empirical study with 17,629 Puppet manifests. We observe TaintPup to have 2.4 times more precision compared to that of SLIC, a state-of-the-art security static analysis tool for Puppet. Our empirical study shows security weaknesses to propagate into 4,457 resources, where a single weakness can propagate into as many as 35 distinct resources. Furthermore, we observe security weaknesses to propagate into a variety of resources, e.g., resources used to manage CI and container-based infrastructure. Our survey-related findings indicate a disconnect between developer perception and empirical characterization of security weakness propagation. Such disconnect further highlights the importance of using TaintPup in Puppet manifest development as it can automatically identify resources that are affected by security weaknesses.

## ACKNOWLEDGMENTS

We thank the PASER group at Auburn University and the Alt-code group at NC State University for their valuable feedback. This research was partially funded by the U.S. National Science Foundation (NSF) award # 2026869, # 2026928, and # 2209636.

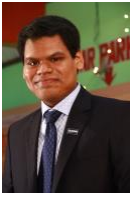

**Akond Rahman** Akond Rahman is an assistant professor at Auburn University. His research interests include DevOps and Secure Software Development. He graduated with a PhD from North Carolina State University, an M.Sc. in Computer Science and Engineering from University of Connecticut, and a B.Sc. in Computer Science and Engineering from Bangladesh University of Engineering and Technology. He won the ACM SIGSOFT Doctoral Symposium Award at ICSE in 2018, the ACM SIGSOFT Distinguished Paper Award at ICSE in 2019, the CSC Distinguished Dissertation Award, and the COE Distinguished Dissertation Award from NC State in 2020. He actively collaborates with industry practitioners from GitHub, WindRiver, and others. To know more about his work visit https://akondrahman.github.io/

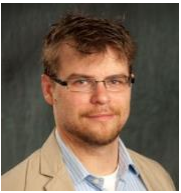

**Chris Parnin** Chris Parnin is an associate professor at North Carolina State University. His research spans the study of software engineering from empirical, human-computer interaction, and cognitive neuroscience perspectives, publishing over 60 papers. He has worked in Human Interactions in Programming groups at Microsoft Research, performed field studies with ABB Research, and has over a decade of professional programming experience in the defense industry. His research has been recognized by the SIGSOFT Distinguished Paper Award at ICSE 2009, Best Paper Nominee at CHI 2010, Best Paper Award at ICPC 2012, IBM HVC Most Influential Paper Award 2013, CRA CCC Blue Sky Idea Award 2016. He research has been featured in hundreds of international news articles, Game Developer's Magazine, Hacker Monthly, and frequently discussed on Hacker News, Reddit, and Slashdot.